\documentclass[twocolumn,aps,prb,graphicx,showpacs]{revtex4}%
\usepackage{amsfonts}
\usepackage{amsmath}
\usepackage{amssymb}
\usepackage{graphicx}%
\begin{document}
\title{On quantum Griffiths effects in metallic systems}
\author{A. J. Millis$^{(1)}$, D. K. Morr$^{(2)}$, J. Schmalian$^{(3)}$}
\affiliation{$^{(1)}$Department of Physics, Columbia University\\
538 West 120th Street, NY, NY, 10027\\
$^{(2)}$Department of Physics, University of Illinois at Chicago, \\
845 W. Tayloer St., Chicago, IL 60607\\
$^{(3)}$Department of Physics and Astronomy and Ames Laboratory,\\
Iowa State University, Ames, IA 50011}

\date{\today}

\begin{abstract}
Elementary analytical extremal statistics arguments are used to analyse the
possibility of quantum Griffiths effects in nearly critical systems with
overdamped dynamics, such as arise in conventional theories of metallic
quantum criticality. The overdamping is found to strongly suppress quantum
tunnelling of rare regions, leading to superparamagnetic rather than quantum
griffiths behavior. Implications for theories of non-fermi-liquid behavior in
heavy fermion materials are discussed.

\end{abstract}

\maketitle

\section{ Introduction}

The interplay of disorder and quantum criticality is a long-standing and still
open problem in condensed matter theory. One aspect of this problem which has
received considerable recent attention is the 'quantum Griffiths' behavior
which has been shown to occur near quantum critical points in certain model
systems \cite{Thill99,McCoy69,Guo96,Rieger96,Fisher98}. The model systems
in which quantum Griffiths behavior has been unambiguously demonstrated all
possess a crucial common feature, namely that in the absence of disorder the
critical degrees of freedom exhibit dissipationless, Hamiltonian spin dynamics
(indeed typically characterized by dynamical exponent $z=1$). However, many
systems of experimental importance involve magnetic degrees of freedom coupled
to conduction electrons \cite{Lohneysen96,Mackenzie01,Lonzarich99,Maple98},
and therefore \textit{overdamped} dynamics implying a pure-system critical
behavior characterized by $z>1$. Extension of the theory of quantum Griffiths
behavior to this case is therefore an important issue. In a series of papers
\cite{Neto98,Neto00,Neto01} Castro-Neto and Jones have argued from various
points of view that such overdamped systems exhibit quantum Griffiths behavior
similar to that exhibited by undamped systems, and they and others have
further argued that this phenomenon is at the heart of the 'non-fermi-liquid'
behavior observed in many heavy fermion materials \cite{Maple98,Stewart01}.

In this paper we examine the issue of quantum Griffiths behavior in nearly
critical systems exhibiting overdamped dynamics, finding that it is
essentially nonexistent, being replaced instead by 'superparamagnetic'
behavior. The essence of our analysis is
this: in undamped models quantum Griffiths effects arise from an interplay
between the low probability of nucleating magnetic 'droplets' in the
paramagnetic state and a low but non-negligible quantum tunnelling of these
droplets. In a metallic, dissipative environment there is a strong suppression
of tunnelling by dissipation, so that the droplets which dominate the
susceptibility behave more or less classically, leading to superparamagnetic
behavior rather than quantum Griffiths behavior.

Our results amount to an implementation of ideas
outlined in \cite{Sherrington75} and to
a generalization, to a non-vanishing density
of defects, of a previously reported analysis \cite{Millis01a} of the
'magnetic droplet' produced by a single, spatially localized defect, and rely
heavily on the results of this previous work. The method used to analyse
the dynamics of a distribution of defects is similar to that
used in \cite{Neto00} and the broad qualitative features of the results
we obtain are very similar to those obtained in that work. 
However, the specifics and the physical implications seem different. The
issue is discussed in more detail in the conclusion.

The outline of this paper is as follows. In section II we present the model
and the method used in our analysis. In section III we show that the approach
reproduces results previously obtained in the dissipationless case. Section IV
presents our new results concerning Griffiths-like behavior in systems with
overdamped dynamics. Section V is a summary, comparison to other work, and
conclusion, and is written so that readers uninterested in the details of the
derivations may obtain from it the essence of our results.

\section{Model and method of solution}

\subsection{Model}

The canonical quantum Griffiths problem concerns the effect of weak disorder
added to a 'pure' (non-disordered) system which possesses an Ising symmetry
and is tuned to be near a quantum critical point. We consider a system in
imaginary time and $3$ spatial dimensions (differences occurring for two
spatial dimensions warrant a separate treatment, which will be presented
elsewhere). The model is described by the action
\begin{equation}
S=S_{\mathrm{static}}+S_{\mathrm{dyn}}+S_{\mathrm{disorder}} \label{S}%
\end{equation}
with
\begin{align}
S_{\mathrm{static}}  &  =\frac{E_{0}}{8\pi}\int_{0}^{\beta}d\tau\int
\frac{d^{3}x}{\xi_{0}^{3}}\left[  \frac{\xi_{0}^{2}}{\xi^{2}}\phi^{2}%
(x,\tau)\right. \nonumber\\
&  \qquad\left.  +\xi_{0}^{2}\left[  \nabla\phi(x,\tau)\right]  ^{2}+\frac
{1}{2}\phi^{4}(x,\tau)\right]  \label{Sstatic}%
\end{align}
Here $\phi$ is a dimensionless scalar order parameter, $E_{0}$ is the basic
energy scale of the theory (perhaps of the order of the mean Kondo temperature
for a heavy fermion system) and is fixed by normalizing the
coefficient of the $\phi^4$ term to unity, 
$\xi_{0}$ is the basic length scale (typically of
the order of a lattice constant), $\xi$ is the magnetic
correlation length, and $\beta$ is the inverse temperature. It is convenient
to define a parameter $r=(\xi_0/\xi)^2>0$ which measures 
distance from criticality. 
We consider 
only parameters such that the 
pure system is in the paramagnetic phase.

We take the disorder to couple to the square of the order parameter via
\begin{equation}
S_{\mathrm{disorder}}=\frac{E_{0}}{8\pi}\int_{0}^{\beta}d\tau\int\frac{d^{3}%
x}{\xi_{0}^{3}}V(x)\phi^{2}(x,\tau) \label{Sdis}%
\end{equation}
and assume it to be Gaussian distributed with correlator ($\left\langle
...\right\rangle $ represents average over configurations of the disorder)
\begin{equation}
\left\langle V(x)V(y)\right\rangle =V_{0}^{2}K\left(  \frac{x-y}{\xi_{0}%
}\right)  \label{VV}%
\end{equation}
where the kernel $K(u)$ decays on the scale $u\sim1$ and satisfies $\int
d^{3}uK(u)=1$. Because we are interested only in length scales $x-y>\xi
_{0}$ we will take $K$ to be a $\delta$ function.
The dimensionless quantity $V_{0}$ parameterizes the strength of the disorder.
Weak disorder corresponds to $V_{0}\ll1$.

The dynamic term $S_{\mathrm{dyn}}$ is crucial to the quantum criticality
described by Eq. \ref{S} and to our subsequent discussions. We consider two
cases: (i) \textit{dissipationless}, $z=1$ \textit{dynamics}, as is the
usually assumed in studies of quantum Griffiths behavior, with
\begin{equation}
S_{\mathrm{dyn}}^{(z=1)}=\frac{E_{0}}{8\pi}\int_{0}^{\beta}d\tau\int
\frac{d^{3}x}{\xi_{0}^{3}}\left(  \frac{\xi_{0}}{c}\right)  ^{2}\left(
\frac{\partial\phi(x,\tau)}{\partial\tau}\right)  ^{2} \label{Sz=1}%
\end{equation}
Here $c$ is a characteristic velocity of the undamped excitations, such that
$c/\xi_{0}$ is an energy presumably of the order of $E_{0}$.\newline(ii)
\textit{Hertz antiferromagnet}, $z=2$ \textit{dynamics}, corresponding to the
generic antiferromagnetic transition in a fermi liquid:
\begin{equation}
S_{\mathrm{dyn}}^{(z=2)}=S_{\mathrm{dyn}}^{(z=1)}+\frac{T}{8 \pi E_{0}}\sum
_{\omega_{n}}\frac{\left\vert \omega_{n}\right\vert }{\Gamma}\int\frac{d^{3}%
x}{\xi_{0}^{3}}\left\vert \phi(r,\omega)\right\vert ^{2} \label{Sz=2}%
\end{equation}
where
\begin{equation}
\phi(r,\omega_{n})=E_{0}\int_{0}^{\beta}\,d\tau\,\phi(r,\tau)\,e^{i\omega
_{n}\tau}%
\end{equation}

In these conventions  the dynamics are {\it dissipative} (i.e. dominated
by the $\Gamma$ term) if 
$\omega < \omega^* \equiv c^2/(\xi^2 \Gamma)$, and non-dissipative at higher
frequencies. One expects in most systems (and finds for example in
a weakly coupled fermi liquid or in the slave boson theory
of the Kondo lattice) that all scales
are of roughly the same order, i.e. that $E_0 \sim c/\xi_0 \sim \Gamma$.

\subsection{Method}

\subsubsection{\textit{Overview}}

The dissipative term in Eq.\ref{Sz=2} corresponds to a long ranged
interaction in time and renders available numerical methods prohibitively
difficult to apply. To analyze the model defined by Eq. \ref{S} we use simple
analytical arguments modelled on those of Ref.~\cite{Thill99}. We note that
the effective dimensionality of the model defined by Eq. \ref{S} is
$d_{\mathrm{eff}}=d+z$. In this paper we consider only the spatial dimension
$d=3$ so we are concerned only with models at and above the upper critical
dimension $d_{c}=4$, so that quantal and thermal fluctuations
of the order parameter in a fixed disorder configuration
can be treated by an essentially mean field 
approximation. The usual fluctuation analysis which justifies the
mean-field approximation for $d_{\mathrm{eff}}>d_{c}$ involves a
translation-invariant model and fluctuations for which momentum is a good
quantum number. Here we must deal with fluctuations in a system whose
translation invariance is broken. These were investigated in
Refs.~\cite{Nayaranan01} and  \cite{Millis01a} 
and were found not to affect the
structure of the static mean field solution when $d_{\mathrm{eff}}\geq d_{c}$
(except for some insignificant changes in some constants).

As noted for example by \cite{Sherrington75},
in the presence of the random potential, the crucial feature of the mean field
solution is the presence of \textit{droplets}: regions in which the order
parameter is locally non-vanishing. Quantum Griffiths effects then arise from
dynamical fluctuations of these droplets; to study them one must
estimate the droplet density and tunnelling rate. We use 
statistical arguments and mean field analysis to estimate
the density and an adaptation 
to the present case of the analysis
presented for a droplet produced by a
single point defect in Ref.~\cite{Millis01a} to estimate the
tunnelling rate.

\subsubsection{\textit{Probability for the existence of a droplet}}

The assumption that the model is at or above its upper critical dimension
means that mean-field theory is a good starting point
\cite{Millis01a,Nayaranan01}. We therefore consider static configurations,
$\phi(x)$, which minimize the combination of Eqs.\ref{Sstatic} and
(\ref{Sdis}). These satisfy
\begin{equation}
\xi_{0}^{2}\nabla^{2}\phi(x)+r\phi(x)+\phi(x)^{3}=-V(x)\phi(x) \label{mfeq}%
\end{equation}
If $V(x)=0$, then because we assume $r>0$ the minimum corresponds to
$\phi(x)=0$; however regions in which $V(x)<0$ can lead to $\phi(x)\neq0$. In
the regions where $V(x)=const.<0$, $\phi(x)$ is roughly constant whereas in
between these regions $\phi(x)$ decays exponentially. We refer to the regions
where $\phi$ is not exponentially small as 'droplets'. If the droplets are
reasonably dilute, one may set $\phi=0$ in the exponential tail regions
\cite{Nayaranan01} and estimate the density of droplets of a given size and
mean amplitude.

To motivate our estimate we first consider solving Eq \ref{mfeq} if
$V(x)=\overline{V}$ for $\left\vert x\right\vert <R$ and $V_{0}=0$ otherwise.
A previous paper \cite{Millis01a} considered a special case of this equation,
with $V_{0}(x)=V\delta^{(d)}(x)$ and the solutions found in that work may
easily be modified for the present case. In $d=3$ one finds that the solution
is, roughly (and neglecting unimportant logarithmic factors in the
$x$-dependence)
\begin{equation}
\phi(x)=\left\{
\begin{array}
[c]{ll}%
\phi_{0}\; & \mathrm{for}\ x<R\\
\frac{\phi_{0}R}{x}e^{-(x-R)/\xi}\; & \mathrm{for}\ R<x\\
\end{array}
\right.  \label{phi3d}%
\end{equation}
In other words, the magnetic order induced by the region of attractive
$V$ is roughly constant inside the region and decays outside it, initially as
$1/x$ and exponentially for distances larger than a correlation length
from the boundary of the attractive potential region.
Inserting the above ansatz, Eq. \ref{phi3d}, into
Eqs. \ref{Sstatic} and \ref{Sdis}, and minimizing the resulting action with
respect to $\phi_{0}$ yields%
\begin{equation}
-\overline{V}=\frac{\xi_{0}^{2}}{\xi^{2}}a(R/\xi)+\phi_{0}^{2}b(R/\xi)
\label{mf3d}%
\end{equation}
with $a(x)=1+3/x+3/x^2$ and $b(x)=1+3/x-4e^{4x}\Gamma[0,x]$. These particular
forms for $a$, $b$ depend on the specific potential configuration studied
(here $\overline{V}=const$ for $x<R$ and $\overline{V}=0$ otherwise)
and on the variational approximation used; but we argue that 
a generic droplet is described by a similar equation with 
$a,b$ functions which vary on the scale $R/\xi\sim1$ and which tend to
unity as $R/\xi\rightarrow\infty$. 
Also in $3$ dimensions $a(x)\sim1/x^{2}$ as
$x\rightarrow0$ while $b(x)$ tends to a constant for $x<1$ \cite{Millis01a}.
The precise forms of $a,b$ affect only nonuniversal details such as widths of
crossover regions. In this paper we shall assume%
\begin{subequations}
\begin{align}
a(x)  &  =1+3x^{-2}\label{a}\\
b(x)  &  =1 \label{b}%
\end{align}
where the $3$ arises from the difference in integrating a constant or
$1/r^{2}$ over $r^{2}dr$.

One sees from Eq \ref{mf3d} that in order to obtain a solution at all the
average potential, $\overline{V}$, must be smaller than a (negative)
$R-$dependent critical value,
\end{subequations}
\begin{equation}
V_{c}=-\frac{\xi_{0}^{2}}{\xi^{2}}a(R/\xi) \label{Vc}%
\end{equation}
which tends to $\xi_{0}^{2}/\xi^{2}$ as $R\rightarrow\infty$ and to a number
of order $1$ as $R\rightarrow\xi_{0}$. As is evident from these formulae, the
natural scale of the droplets is the correlation length $\xi$ which diverges
as the quantum critical point is approached.

Eqs. \ref{mf3d} and \ref{Vc} thus imply  
that one obtains a droplet in a region of linear dimension
$R$ only if the average value $\overline{V}$
of the potential in that region is larger than
a value of the order of
$V_{c}(R/\xi)$ ($V_c$ is not an exact estimate because
it pertains to the idealized disorder configuration
discussed above). The standard estimate of the probability of
a region of linear dimension $R$ with mean potential $\overline{V}$ is
\begin{equation}
P(R^{3},\overline{V})\sim\frac{\left(  R/\xi_{0}\right)  ^{3/2}}{\sqrt{\pi}V_{0}}%
\exp\left(  -\left(  \frac{R}{\xi_{0}}\right)  ^{3}\left(  \frac{\overline{V}}{V_{0}%
}\right)  ^{2}\right)  \label{Pave}%
\end{equation}
and we therefore argue that the density $N(R^{3},\phi_{0}^{2})$ 
of droplets of amplitude $\phi_{0}^{2}$ 
and core size $R$, must be proportional to
$\frac{1}{V_{0}}\exp\left(  -\frac{R^{3}\left(  \phi_{0}^{2}+V_{c}%
(R/\xi)\right)  }{V_{0}^{2}}^{2}\right)  $. This argument does
not determine the preexponential factors (which involve, e.g. the
issue of whether the region of size $R$ considered in Eq. \ref{Pave}
is part of a larger region which can sustain a droplet
and numerical factors arising from the difference
between idealized disorder configuration and typical one,
which we have absorbed into $V_0$ and $V_c$ ).
Because some of our subsequent
considerations will require an estimate of the preexponential factors,
we present the  following arguments to fix them.

We begin by making a rough estimate of the fraction of sites contained in
droplets (i.e. of the fraction of sites
having a $\phi_0^2>0$), as a function of distance from
criticality. As noted above, 
in principle within mean field theory $\phi_0^2$ 
is non-vanishing everywhere, but we neglect the regions where
it is exponentially small, in other words we set $\phi_0=0$ in the 
'inter-droplet' regions.
To perform the estimate we coarse-grain 
the theory to the scale $\xi$. A given
correlation volume $\xi^{3}$ will have a non-vanishing $\phi_{0}^{2}$ if the
potential averaged over the droplet volume, $V_{\mathrm{ave}},$ is larger than
$V_{c}(\xi)\approx a(1)\xi_{0}^{2}/\xi^{2}$. From Eq \ref{Pave} we see that
the probability $P_{\phi}$ that a given correlation volume will have a
non-vanishing $\phi_0$, i.e. a
$V<V_{c}(\xi) $, is (recall $V_c<0$)
\begin{equation}
P_{\phi}=\frac{1}{2}\left(  1-\operatorname{erf}\left[  \left(  \frac{\xi}%
{\xi_{0}}\right)  ^{3/2}\frac{V_{c}(\xi)}{V_{0}}\right]  \right)  \label{Pphi}%
\end{equation}
where $erf$ is the error function.

Clearly, a picture of independent droplets must break down if $P_{\phi}$
exceeds the percolation probability $P_{\mathrm{perc}}$ at which the 
set of correlation volumes with
non-vanishing $\phi_{0}$ percolate. Use of Eq. \ref{Pphi} \ and the
estimate for three dimensional cubic lattices $P_{\mathrm{perc}}\approx0.2$
shows that percolation will have occurred by the time $\xi$ exceeds
$\xi_{\mathrm{perc}}\approx2.8a(1)^{2}/V_{0}^{2}$. ($\xi_{\mathrm{perc}}$ is
an underestimate because droplets larger than $\xi$ may occur). These
estimates also show that the natural scale  for $\xi$ is $V_{0}^{-2}$ and
strongly suggest that the probability that 
a given site is in a droplet (of any
size) is a function only of the combination $\xi V_{0}^{2}$.

We therefore argue that the prefactors in the droplet density must be such
that the total probability of finding a site in a droplet, $P_{\mathrm{tot}%
}=\xi_{0}^{-6}\int dR^{3}d\phi_{0}^{2}R^{3}N(R^{3},\phi_{0}^{2})$ must be a
function only of $\xi V_{0}^{2}$ and must be of the order of $P_{\mathrm{perc}%
}$ when $\xi$ is of the order of $\xi_{\mathrm{perc}}$. This implies
\begin{equation}
N[R^{3},\phi_{0}^{2}]\ =\frac{R^{-9/2}}{C_{V_{0}}V_{0}}\exp\left(
-\frac{R^{3}\left(  \phi_{0}^{2}+V_{c}(R/\xi)\right)  }{V_{0}^{2}}^{2}\right)
\label{P3d}%
\end{equation}
where the factor of $R^{-9/2}$ ensures the correct scaling with $\xi$ and the
numerical factor $C_{V_{0}}\approx11.25$ ensures that when $\xi=\xi
_{\mathrm{perc}}$, we have $P=P_{\mathrm{perc}}$. We emphasize that these
formulae are phenomenological and must in particular break down when $\xi$
approaches $\xi_{\mathrm{perc}}$.

It is convenient to adopt a dimensionless system of units in which%
\begin{align}
R  &  =y\xi\label{ydef}\\
\phi_{0}  &  =f\frac{\xi_{0}}{\xi}\label{fdef}\\
\xi &  =u\frac{\xi_{0}}{V_{0}^{2}} \label{udef}%
\end{align}
for which
\begin{align}
&  N(y^{3},f^{2})dy^{3}df^{2}=\frac{y^{-9/2}}{C_{V_{0}}u^{1/2}\xi^{3}%
}\nonumber\\
&  \times\exp\left(  -\frac{y^{3}\left(  f^{2}+a(y)\right)  ^{2}}{u}\right)
dy^{3}df^{2} \label{pscaled}%
\end{align}

The factor of $\xi^{-3}$ expresses the fact that if the probability of a given
site being in a droplet is a function only of $u$, then the density of
droplets must be smaller by an extra factor of the typical droplet volume
$\xi^{3}$.

\subsubsection{Tunnelling of the droplet for undamped, $z=1$, dynamics}

We now estimate the rate $\omega_{\mathrm{tun}}$ at which a droplet
characterized by the mean amplitude $\phi_{0}$ and length scale $R$ tunnels in
the case of undamped, $z=1$ dynamics by performing a variational instanton
calculation using Eqs.\ref{Sstatic}, \ref{Sz=1} and the solution
Eq.~\ref{phi3d}. In the simplest estimate one assumes that the droplet
maintains its shape while collapsing and re-forming. To estimate the action
associated with this process we write the droplet solution as
\begin{equation}
\phi(x,\tau)=\phi(x)\eta(\tau) \label{ansatz}%
\end{equation}
Substitution into Eqs. \ref{Sstatic}, \ref{Sz=1} leads to
\begin{equation}
S_{\mathrm{inst}}=S_{\mathrm{kin}}+S_{\mathrm{barrier}} \label{Sinst}%
\end{equation}
$S_{\mathrm{kin}}$ involves the integral of $\left(  \partial_{t}\phi\right)
^{2}$ over the droplet and as noted in Ref.~\cite{Millis01a} involves the
$1/r$ 'tail' of the droplet in a crucial manner; in contrast, the cost
$S_{\mathrm{barrier}}$ of creating the instanton does not. One obtains
\begin{align}
S_{\mathrm{kin}}^{(z=1)}  &  =C_{\mathrm{kin}}\xi f^{2}y^{3}a^{\prime}%
(y)\int\frac{d\tau}{E_{0}}\left(  \frac{\partial\eta}{\partial\tau}\right)
^{2}\label{Skinz1}\\
S_{\mathrm{barrier}}  &  =C_{\mathrm{barrier}}\xi^{-1}f^{4}y^{3}b^{\prime
}(y)\nonumber\\
&  \times\int E_{0}d\tau\left(  -2\eta(\tau)^{2}+\eta(\tau)^{4}\right)
\label{Sbarrier2}%
\end{align}
Here $C_{\mathrm{kin}}$ and $C_{\mathrm{barrier}}$ are nonuniversal constants.
$C_{\mathrm{kin}}$ involves the square of the ratio $E_{0}/(c/\xi_{0})$ of the
basic energy scale to the kinetic (or zone boundary magnon) energy while
$C_{\mathrm{barrier}}$ is just a number.  In the
approximation we have employed $C_{\mathrm{kin}}=E_0^2\xi_0^2/c^2$ and
$C_{\mathrm{barrier}}=1$. The functions
$a^{\prime}=\int x^2dx \phi_0(x)^2$ and 
$b^{\prime}=\int x^2dx \phi_0(x)^4$ are 
functions with behavior similar to $a,b$; in
our explicit calculations we set $a^{\prime}=a/3=1/3+y^{-2}$ and 
$b^{\prime}=b/3=1/3$ for simplicity; again different choices 
affect only nonuniversal details.

The action associated with one instanton may now be determined by a standard
minimization of Eqs \ref{Skinz1},\ref{Sbarrier2} and is
\begin{equation}
S_{\mathrm{inst}}^{(z=1)}=S_{1}d(y)f^{3}y^{3} \label{Sinstz1}%
\end{equation}
For the present model in the present approximation the nonuniversal constant
$S_{1}=\sqrt{C_{\mathrm{kin}}C_{\mathrm{barrier}}/3}$and $d(y)=3\sqrt{a^{\prime
}(y)b^{\prime}(y)}=\sqrt{a(y)}$ where the last equality follows from our
simplifying assumptions $a'=3a$ and $b'=3b$. The value of $S_{1}%
$controls the width of the crossover regime before the universal behavior is
reached, and is linearly proportional to
$E_{0}\xi/c$ 

The tunnelling rate is then given by
\begin{equation}
\omega_{\mathrm{tun},z=1}=\omega_{0}e^{-S_{\mathrm{inst}}(\tau_{0})}\ .
\label{wtunz1}%
\end{equation}
Here, $\omega_{0}$ is an attempt frequency presumably of order $E_{0}$ whose
value is beyond the scope of this theory.

To conclude this section we briefly estimate the action associated with a
different tunnelling mechanism, namely nucleation of a domain wall. For small
droplets ('core size' $R$ less than $\xi$) the important process was
shown to be collapse and reformation of the entire droplet \cite{Millis01a}. 
We therefore need consider only the case
$R\gg\xi.$ We observe that by expanding about the static uniform solution one
obtains a domain wall with width $W\sim\phi_{0}^{-1}.$ The kinetic term
associated with the domain wall motion therefore has one fewer factor of the
small quantity $\phi_{0}\sim f/\xi \sim V_0^2f$, 
leading to a larger action and hence a
smaller rate, in the weak disorder, near criticality limit. We note in passing
that for $\phi_{0}\sim1$ the powers of $R$ will be the same as we have
considered but the extra factor of $\phi_{0}^{-1}$ will work in the other
direction, favoring domain wall motion.

\subsubsection{Tunnelling of the droplet for overdamped, $z=2$, dynamics}

For overdamped dynamics two important differences occur. First, as shown in
Refs.~\cite{Leggett87,Neto00,Millis01a} the damping changes the action associated
with a single instanton, strongly suppressing the \textit{bare} tunnelling
rate relative to that found for undamped dynamics. Essentially, the tunnelling
is limited by the droplet's ability to move through a viscous medium rather
than by its ability to climb over a barrier. Second, and much more important,
the overdamped dynamics leads to a long-ranged (in time) instanton-instanton
interaction, which reduces the tunnelling rate further and indeed drives it to
zero if the damping exceeds a critical value, as noted by previous authors
\cite{Neto00,Millis01a}.

To calculate the effects of damping we insert the ansatz, Eq. \ref{ansatz}
into Eq \ref{Sz=2}. The new term arising from the overdamped dynamics is%
\begin{equation}
S_{\mathrm{diss}}=\frac{\gamma}{4}\int d\tau d\tau^{\prime}\frac{d\eta}{d\tau}%
\frac{d\eta}{d\tau^{\prime}}\ln(\frac{(\tau-\tau^{\prime})^2+\tau_m^2}{\tau_{m}^2%
})\label{sdiss}%
\end{equation}
with $\tau_{m}$ a 'microscopic' time of the order of $\omega^*=c^2/(\xi_0^2\Gamma$. 
The net
dissipative coefficient $\gamma$ is given for the Hertz antiferromagnet by
\cite{Millis01a}
\begin{equation}
\gamma=\frac{E_{0}}{4\pi\Gamma}\int\frac{d^{3}x}{\xi_{0}^{3}}\phi_{0}%
(x)^{2}=c_{\gamma}f^{2}y^{3}a^{\prime\prime}(y)\xi/\xi_{0}\label{gamma}%
\end{equation}
The approximations employed in the previous section
imply that the nonuniversal constant $c_{\gamma}=E_{0}/\Gamma$ 
and $a^{\prime\prime}=a^{\prime}$.
In  a generic system one expects all scales to be of roughly
the same order, so that in particular $c_{\gamma}$ is expected
to be of order unity.

The estimate of $\gamma$ is subject to the important caveat that the electron
bath which causes the dissipation can penetrate the entire droplet. A
reasonable estimate of the penetration depth, $L_{p}$, may be obtained by
dividing the electron velocity, $v_{\mathrm{F}}$, by the magnitude of the
order parameter; in rescaled units $L_{p}/\xi\sim v_{\mathrm{F}}/(E_{0}f)$. We
shall see below Eq. \ref{flimit}
that the parameters are such that the electrons can penetrate
the entire droplet.

We have not been able to solve analytically for the instanton; instead we
estimate the action by inserting the variational ansatz%
\begin{equation}
\frac{d\eta}{d\tau}=2\frac{\Theta(\tau_{0}^{2}-4\tau^{2})}{\tau_{0}%
}\label{z2ansatz}%
\end{equation}
into Eqs \ref{Sz=2},\ref{Sbarrier2} obtaining $S=S_{kin}+S_{diss}+S_{barrier}$
with

\begin{equation}
S_{kin}=\frac{2C_{kin}\xi y^3 a'(y) f^2}{E_0 \tau_0 \xi_0}
\label{Skinz2}
\end{equation}

\begin{equation}
S_{diss}=2 c_{\gamma} \frac{\xi}{\xi_0} y^3 a'(y) f^2 ln(c_d \tau_0/\tau_m)
\label{Sdissz2}
\end{equation}

\begin{equation}
S_{barrier}=\frac{2}{15} C_{barrier} \frac{\xi_0}{\xi} y^3 b'(y) f^4 E_0\tau_0
\label{Sbarrierz2}
\end{equation}

where  $ln(c_d)=\int_{-1/2}{1/2}dxdyln(1+(x-y)^2) \approx 0.1152...$.

Minimization over the instanton duration then leads to

\begin{equation}
1= \frac{c_{\gamma}}{C_{kin}}\tau_0 E_0 +
\frac{C_{barrier}}{15 C_{kin}}\frac{b'f^2 \xi_0^2}{a' \xi^2}(\xi_0/\xi)^2 
(E_0 \tau_0)^2
\label{tauz2eq}
\end{equation}

As previously remarked, we expect the ratios of 
the various dimensional parameters
to be of the order of unity; also as we shall see below,
in this problem
the important droplets have
$f \sim \xi^{-1/2}$, so that provided $\Gamma$
is less than a number of the order unity times 
$\frac{\xi c}{f \xi_0} \sim \xi^{3/2}$
(within our approximations the precise numerical factor is $\sqrt{15}$)
the $\tau_0^2$ term is negligible
and one has

\begin{equation}
\tau_{0}=\frac{\Gamma \xi_0^2}{c^2}
\label{tauz2}%
\end{equation}
and thus 

\begin{equation}
S_{\mathrm{inst}}^{(z=2)}=c_{\gamma}C_2f^{2}\xi\xi_{0}^{-1}y^{3}a(y)%
\label{Sinstz2}%
\end{equation}
where $C_2$ (=2.283... in the present approximations)
is a numerical factor of the order of unity arising
from combining the factors in Eqs \ref{Skinz2} and \ref{Sdissz2}).

We observe that for the value of $\tau_{0}$ given in Eq \ref{tauz2}, the term
written in Eq \ref{Sinstz2} is  larger than $S_{\mathrm{barrier}}$ (Eq
\ref{Sbarrier2}) by two powers of the correlation length
(provided that the quantity $f$ is of order unity or less,
as is the case for the situations considered here.)
Thus,  in the metallic case and near to criticality, the 
difficulty in tunnelling arises from
moving through the viscous medium, not climbing over the barrier.
This result was noted previously \cite{Millis01a}.

The bare tunnelling amplitude is thus
\begin{equation}
\omega_{\mathrm{bare}}^{(z=2)}=\omega_{0}e^{-S_{\mathrm{inst}}^{(z=2)}}
\label{wbared3z2}%
\end{equation}
and is much smaller than in the dissipationless case, because 
of the factor $f\xi$ in the argument of the exponential.

The standard macroscopic quantum tunnelling arguments
\cite{Leggett87,Neto00,Millis01a} imply that 
the instanton-instanton interaction
renormalizes the bare tunnelling rate so that if $\gamma<1$ then
the $T=0$ tunnelling rate is
\begin{equation}
\omega_{\mathrm{tun}}=\omega_{0}\left(  \frac{\omega_{\mathrm{bare}}}%
{\omega_{0}}\right)  ^{\frac{1}{1-\gamma}} \label{wtundiss}%
\end{equation}
whereas if $\gamma>1$ tunnelling stops at $T=0$.
We see from Eq. \ref{gamma} that
$\gamma$ is a strong function of the droplet size and amplitude; droplets
which may tunnel (i.e have $\gamma<1$) have a very weak amplitude even in
rescaled units: $f\sim\xi^{-1/2}$.

Eq \ref{wtundiss} is a zero temperature result. At $T>0$ the 
'Caldeira-Leggett' renormalization is temperature dependent.
The key question for this paper is the temperature
at which  $\omega_{tun}(T)<T$. If $\gamma>1$ then 
$\omega_{tun}(T)<T$ at all $T<E_0$, implying that the droplet
behaves classically at all $T$. If $\gamma<1$ then 
the usual arguments shows that $\omega_{tun}(T)$ drops
below $T$ when $T$ becomes greater than $\omega_{tun}(T=0)$,
so that Eq \ref{wtundiss} gives the temperature
scale separating a high-T region, in which the droplet behaves
classically, from the low-T region, in which it
behaves quantum mechanically.

\section{Estimate of Quantum Griffiths Behavior}

\subsection{Overview}

The standard Griffiths estimate is that a droplet of magnetic moment
$M_{d}=\int d^{3}r\,e^{i\overrightarrow{Q}\cdot\overrightarrow{r}}\phi_{0}(r)$
($\overrightarrow{Q}$ is the ordering vector) and tunnelling frequency
$\omega_{\mathrm{tun}}[R,\phi_{0}^{2}]$ gives rise to a susceptibility
$\chi_{d}$ proportional to $M_{d}^{2}/\left(  \omega_{\mathrm{tun}}+T\right)
$. The susceptibility of a system with a distribution of droplets is then
given by
\begin{equation}
\chi(T)=\int d^{3}R\,d^{2}\phi_{0}\,\frac{N(R^{3},\phi_{0}^{2})\,M_{d}%
^{2}[R,\phi_{0}]}{\omega_{\mathrm{tun}}[R,\phi_{0}^{2}]+T}%
\end{equation}
For a droplet in an antiferromagnetic system, we find $M_{d}$ is a random
function with magnitude $\phi_{0}R$--the term proportional to $R$ comes from
the boundary of the droplet, where the order parameter amplitude is dropping
and the cancellation over one unit cell of the antiferromagnetic order is not
complete. A different dependence would change prefactors but not affect our
results crucially.

It is convenient to introduce an explicit integral over frequency, writing%
\begin{equation}
\chi(T)=\xi^{-3}\int d\omega\,\frac{I(\omega)}{\omega+T}\label{chi2}%
\end{equation}
so that after conversion to dimensionless units we have
\begin{equation}
I(\omega)=\int dy^{3}\,df\,^{2}(\xi^3N(y^{3},f^{2}))f^{2}y^{2}\delta(\omega
-\omega_{\mathrm{tun}}(y,f))\label{I}%
\end{equation}

The prefactor $\xi^{-3}$ in $\chi$
arises because each droplet has magnetic
moment of the order of unity and the density of droplets is
$\xi^{-3}$. The quantity $\xi^3N$ has no explicit dependence on
$\xi$ (see Eq. \ref{pscaled}).

We will use the delta function to eliminate the $f$ integral in $I$ and
perform the integration over y either numerically or via an extremal value argument.

\subsection{z=1}

Using Eq. \ref{wtunz1} yields
\begin{equation}
f(\omega,y)=\left(  \frac{\ln\left(  \frac{\omega_{0}}{\omega}\right)  }%
{S_{1}d(y)}\right)  ^{1/3}\frac{1}{y}\label{phi0z1}%
\end{equation}
Substituting this result into Eq.\ref{I} yields%
\begin{equation}
I(\omega)=\frac{2\ln^{1/3}(\omega_{0}/\omega)}{\omega S_{1}^{4/3}}\int
_{0}^{\infty}\frac{\xi^3N(y^{3},f^{2}(\omega,y))dy}{d(y)^{4/3}}\label{Iz=1}%
\end{equation}
where $N(y,f(\omega,y))$ is $N(y,f)$ \ (Eq \ref{P3d}) with $f$ given by Eq
\ref{phi0z1}.

In the limit of very low frequency one may use asymptotic methods to analyse
the integral in Eq \ref{Iz=1}; the extremum is at
\begin{equation}
y_{\max}=\frac{\ln^{1/3}\left(  \frac{\omega_{0}}{\omega}\right)  }{\sqrt
{3}S_{1}^{1/3}} \label{ymaxz1}%
\end{equation}
Substitution leads to
\begin{equation}
\chi(T)\sim\frac{1}{\xi^{3}C_{V_{0}}}\frac{1}{T^{1-d_{\mathrm{asympt}}%
}} \label{chiz1}%
\end{equation}
with (restoring units)
\begin{equation}
d_{\mathrm{asymp}}(\xi)=\frac{16}{3\sqrt{3}S_{1}}\frac{1}{\xi V_{0}^{2}}
\label{dasymp}%
\end{equation}

This is the familiar quantum Griffiths result: if one is sufficiently close to
the pure system critical point ($d(\xi)<1$) then the susceptibility diverges,
with degree of divergence characterized by an exponent which approaches unity
proportional to one power of the inverse correlation length. 

Note that the
prefactor in Eq \ref{chiz1} rapidly
vanishes as criticality is approached, so although the
susceptibility diverges more strongly, the amplitude of
the divergence decreases.  Note further that  
in the asymptotic limit, $f\approx 1$ so that the mean
order parameter density (integrated order parameter divided by droplet volume)
is of the order of $\xi^{-1}$. Thus the picture that emerges is of large, 
weak droplets.

\begin{figure}[ht]
\includegraphics[width=3.0in]{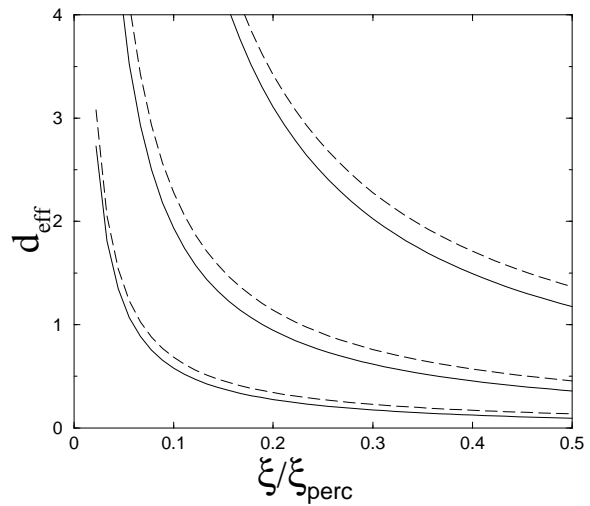}
\caption{Solid lines: calculated effective 
Griffiths exponent for undamped ($z=1$) case
defined in Eq. \ref{deff} plotted vs correlation length normalized to
correlation length $\xi_{perc}$ at which droplets percolate , with (from top
to bottom) non-universal coefficient $S_{1}=.1,.3,1$ and frequency 
$\omega=10^{-3}\omega_0$. Dashed lines: asymptotic result (Eq \ref{dasymp})
for same parameters.}
\end{figure}

We have evaluated 
\begin{equation}
d_{\mathrm{eff}}(\omega)=1+\frac{d\ln\left[  I(\omega)\right]  }{d\ln(\omega)}
\label{deff}%
\end{equation}
via a numerical computation of Eq.\ref{Iz=1}.
Fig. 1 shows $d_{\mathrm{eff}}(\xi,\omega)$ as a function of $\xi$ for
$\omega=0.001\omega_{0}$ and several different values of the non-universal
parameter $S_{1}$ (solid lines) along with the asymptotic limit
estimates\ from Eq \ref{dasymp}. We observe that for these low frequencies and
not too long $\xi$ the asymptotic limit provides a reasonable 
(but not perfect) estimate of the
effective exponent: relative corrections are of the order of 
$(\xi/ ln(\omega_0/\omega)^{2/3})$. 
We see also  that depending on the value of the
non-universal parameter $S_{1}$, the effective exponent may remain above the
critical value of unity (corresponding to a non-divergent susceptibility)
until $\xi$ becomes of the order of $\xi_{perc}$. For $\xi$ of 
the order of $\xi_{perc}$ the standard quantum griffiths
approximation (independent droplets) breaks down, and one must deal instead
with the critical singularities appropriate to a phase transition
in a disordered system; in other words with the still unsolved
problem of the mixing of quantum critical and quantum griffiths singularities.

\subsection{z=2}

For overdamped dynamics, some droplets will have $\gamma>1$ and therefore will
not tunnel at all at $T=0.$ The function $I(\omega)$ will thus have a
contribution proportional to $\delta(\omega)$ leading to the $1/T$ behavior
expected of classical droplets. For those droplets which do tunnel we must use
Eq \ref{wtundiss} in Eq \ref{I}. We write
\begin{equation}
I(\omega)=I_{0}\delta(\omega)+I_{rest}(\omega) \label{Idiss}%
\end{equation}
with $I_{rest}$ given by Eq \ref{I} and $I_{0}$ by
\begin{equation}
I_{0}=\int dy^{3}df^{2}f^{2}y^{2}N(y^{3},f^{2})\Theta(\gamma(y,f)-1)
\label{i0}%
\end{equation}
From Eq \ref{chi2} we see that if $I_{0}$ is appreciable, then $\chi\sim1/T$:
this is the superparamagnetism expected from essentially classical droplets.

We begin by estimating $I_{0}$. The $\Theta$ function limits the $f$
integration to
\begin{equation}
f^{2}>f_{\min}^{2}(y)=\frac{\xi_{0}}{c_{\gamma}\xi y^{3}a(y)} \label{flimit}%
\end{equation}
Note that for large $\xi$, $f_{\min}\ll a(y)$. Further, the typical 
scale for $f$ is $\xi^{-1/2}$ so that the penetration depth 
$L_p$ of electrons into the droplet is large: $L_p/\xi \sim \xi^{1/2}$
so the assumption that electrons penetrate the droplet is indeed
self consistent.
\begin{figure}[ht]
\includegraphics[width=3.0in]{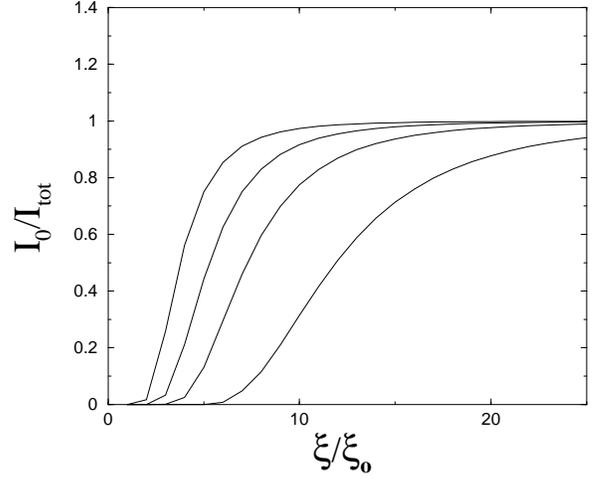}
\caption{Ratio of density of magnetization of non-tunnelling droplets $I_{0}$
(Eq \ref{i0}) to total density of droplets $I_{tot}=\int d\omega I(\omega)$
for overdamped case and non-universal constant $c_{\gamma}=0.1$ (larger values
of $c_{\gamma}$ lead to an $I_{0}/I_{tot}\approx1$ even for much smaller
values of $\xi$, as a function of correlation length \textit{(not normalized
to disorder strength)} for dimensionless disorder
strength $V_{0}=1$, (top curve) $.7,.5,.3$. 
Note that for all reasonable parameters a non-negligible
fraction of droplets do not tunnel at all.}
\end{figure}

Use of Eq \ref{flimit} in Eq \ref{i0} gives%

\begin{align}
I_{0}(\xi) &  =\frac{3\sqrt{\pi}}{2C_{V_{0}}}\int_{0}^{\infty}%
y^{-2}dy\left[  \frac{\sqrt{u}e^{-y^{3}(f_{\min}^{2}(y)+a(y))^{2}/u}}%
{\sqrt{\pi}y^{3/2}}\right.  \nonumber\\
&  \left.  +a(y)\left(  \operatorname{erf}\left(  \frac{y^{3/2}\left(
f_{\min}^{2}(y)+a(y)\right)  }{u^{1/2}}\right)  -1\right)  \right]
\label{iofin}%
\end{align}
$I_{0}$, normalized to the total weight in $I$, $\int d\omega I(\omega)$ is
plotted in Fig. 2 as a function of $\xi$ $\ $\ for different values of the
disorder strength $V_{0}$. We see that the factor of $\xi^{-1}$ in Eq
\ref{flimit} means that  as criticality is
approached, almost all of the weight in the droplet probability distribution
is in droplets which do not tunnel. 

For the droplets which are able to tunnel at frequency $\omega$, we find from
Eqs. \ref{Sinstz2} and \ref{wtundiss} that
\begin{equation}
f_{z=2}^{2}(y)=\frac{\xi_{0}}{c_{\gamma}\xi y^{3}a(y)}\frac{\ln\left(
\frac{\omega_{0}}{\omega}\right)  }{C_{2}+\ln\left(  \frac{\omega_{0}}{\omega
}\right)  }%
\end{equation}
Note that in contrast the expression for $f$ in the $z=1$-case shown in
Eq. \ref{phi0z1}, in the $z=2$ case, $f$ does \textit{not} diverge as
$\omega\rightarrow0$. As in the $z=1$ case considered above, one obtains an
expression for $I_{\mathrm{rest}}(\omega)$ by substituting the result for $f$
into Eq \ref{I} yielding%

\begin{align}
I_{\mathrm{rest}}(\omega)  &  =\frac{3\xi_{0}^{3}}{\omega C_{V} c_{\gamma}^3
\xi^{3}%
}\frac{C_{2}\ln\left(  \frac{\omega_{0}}{\omega}\right)  }{\left(  C_{2}%
+\ln\left(  \frac{\omega_{0}}{\omega}\right)  \right)  ^{3}}\nonumber\\
&  \times\int\frac{dyy^{-1/2}e^{-\frac{y^{3}\left(  f_{z=2}^{2}%
(y)+a(y)\right)  ^{2}}{u}}}{\left(  y^{3}a(y)\right)  ^{2}}
\label{irest2}%
\end{align}
The resulting expression is to good accuracy proportional to $1/\omega$ times
logarithms. The physics is that even the average of droplets which are able to
tunnel is dominated by those droplets on the verge of freezing, leading again
to a superparamagnetic contribution to the susceptibility.

We have numerically evaluated the integral in Eq. \ref{irest2} for parameters
such that $I_{0}$ is not too large. Sample results are shown in Fig.~3,
which plots the quantity $J_{rest}=\omega I_{rest}$ for
a relatively {\it small} value of the damping.  
The frequency dependence is a consequence of the
logarithmic factors in Eq \ref{irest2}; the non-vanishing intercept as
$\omega\rightarrow0$ means that up to logarithmic corrections the
contribution to the
susceptibility arising from this term is $\sim 1/T$.
\begin{figure}[ht]
\includegraphics[width=3.0in]{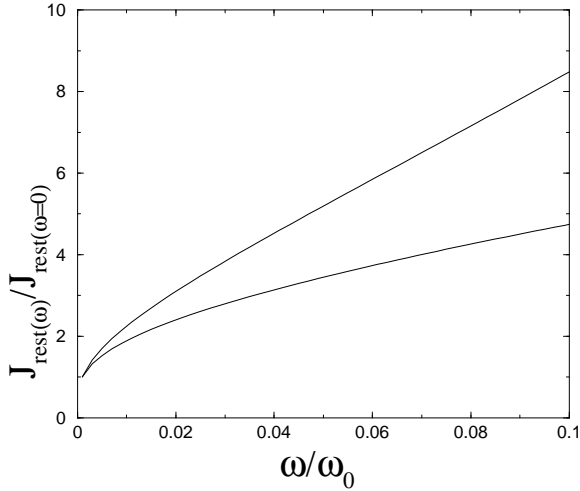}
\caption{ Contribution $J_{rest}(\omega)=\omega I_{rest}(\omega)$ 
(Eq \ref{irest2}) of tunnelling droplets to
susceptibility integral, plotted vs frequency for non-universal constants
$V_0=0.5$, $C_{2}=1$ and relatively
small value of damping coefficient
$c_{\gamma}=0.1$ at $\xi=5$ (top curve) and $\xi=20$
(bottom curve).
}
\end{figure}

\section{Conclusion}

This paper presents an investigation of the possibility of quantum Griffiths
effects in three dimensional metallic system which, when 'pure'
(non-disordered) is near an antiferromagnetic quantum
critical point with Ising symmetry. For comparison we present also a parallel
investigation of quantum Griffiths effects in a model of an insulating system
near a similar critical point. The key feature of metallic systems is the
dissipative dynamics arising from the particle-hole continuum of electrons; in
the model insulating system the dynamics are undamped. Comparison of the two
calculations shows that dissipation suppresses quantum Griffiths effects
completely, leaving instead an effectively superparamagnetic behavior.

A simple precis of our results follows. Quantum Griffiths effects are a
consequence of randomness: essentially, in a random system which is on average
in the paramagnetic phase, regions ('droplets') may occur in which the
randomness pushes the system locally to the ordered side of the phase diagram,
so that local formation of an order parameter is favored. In certain
circumstances (first noted by McCoy \cite{McCoy69}) these droplets may
dominate the response. In this situation one may approximately write the
susceptibility, $\chi$, as an average over droplets times a susceptibiltiy for
each droplet, i.e.
\begin{equation}
\chi=\int_{\mathrm{droplets}}\,P(\mathrm{droplet})\,\chi_{\mathrm{droplet}}%
\end{equation}
We have used simple extremal statistics arguments (similar to those used
by Thill and Huse \cite{Thill99}) to estimate the droplet probability
distribution $P(\mathrm{droplet})$ and an extension of earlier work which
studied a particular class of droplets \cite{Millis01a} to obtain the
susceptibility $\chi_{\mathrm{droplet}}$ of a given droplet. We were then able
to perform the average over droplets and obtain an estimate for the susceptibility.

This method reproduces the essential features of the standard results for
quantum Griffiths effects in undamped (insulating) systems, namely that the
low $T$ behavior of the susceptibility is governed by a new exponent
$d_{\mathrm{eff}}$ given by the product of the inverse correlation length
$\xi^{-1}$ and inverse mean square disorder amplitude $V_{0}^{-2}$ and a
non-universal number (which we estimate for the particular model we consider).
A divergent susceptibility results when $d_{\mathrm{eff}}$ becomes less than
unity, and the results are functions only of $\xi V_{0}^{2}$. \ We note one
additional interesting finding. The standard arguments which produce the
standard quantum Griffiths results are based on a picture of dilute 'droplets'
and apply only if the $\xi$ is not too large (otherwise the droplets
percolate, and an isolated droplets picture fails). For the model we consider
we obtain an estimate for the critical value of $\xi$, and find that depending
on the value of the non-universal factor in $d_{\mathrm{eff}}$, droplets may
reach the percolation point before the Griffiths exponent drops below unity.
In other words, in the models we consider the existence of a quantum Griffiths regime 
(which one may somewhat imprecisely define as a divergent susceptibilty
arising from fluctuations of isolated droplets) is not guaranteed--it may or
may not occur depending on the value of a non-universal coefficient. 
Sufficiently near a critical point a
regime of divergent susceptibility does of course occur, but the proper
theory of this regime would have to go beyond the model of isolated droplets
and treat correctly the mixing of critical and griffiths singularities.

We also found that for undamped systems near antiferromagnetic
critical points the {\it amplitude} of the divergent term in
the susceptibility vanishes rapidly as criticality is approached,
indeed as $\xi^{-3}$, essentially because each relevant droplet
has a magnetic moment of the order of unity and as criticality
is approached the droplets get larger in size but fewer in number.

The main new result of our work, however, pertains to metallic systems with
overdamped (dissipative) dynamics. For these systems (i.e. for quantum
critical phenomena in metals) the answer is entirely different. The physics in
the undamped case is a balance between the probability of a droplet occurring
(which vanishes rapidly as the droplet size or amplitude increases) and
$\chi_{\mathrm{droplet}}$, which is of the order of the inverse of the quantum
tunnelling rate of the droplet and diverges rapidly as the droplet size or
amplitude increases. The effect of dissipation is to strongly decrease the
tunnelling rate, and indeed to drive it to zero for droplets larger than a
particular, amplitude-dependent, size. For relevant parameters we find
that a non-vanishing density of droplets does not tunnel at $T=0$;
these give rise to a 'superparamagnetic' ($\chi \sim 1/T$)
susceptibility rather than a quantum griffiths (continuously varying
exponent) behavior. For those droplets which do behave quantum
mechanically, the effect of dissipation on the tunneling
rate is found to change  the balance
between probability and $\chi_{\mathrm{droplet}}$ dramatically. We find that 
even considering only the droplets which can tunnel quantum
mechanically, those
which dominate the integral for $\chi$ are those which are right on
the edge of classical (non-tunnelling) behavior,
leading again to superparamagnetism rather than to quantum Griffiths behavior.
We also find that the dependence on parameters is different: in the undamped
case, apart from prefactors the mean square disorder strength $V_{0}^{2}$ and
the correlation length enter via the combination $\xi V_{0}^{2}$. In the
damped case additional factors of $\xi$ occur which drive the system more
rapidly to classical behavior.

Our results raise questions
about  the claims \cite{Maple98,Neto98,Neto00,Neto01} that
quantum Griffiths effects are important in heavy fermion materials, which are
precisely three dimensional metals with Ising symmetry, typically near
antiferromagnetic quantum critical points. Ref.~\cite{Maple98} contains a
phenomenological description of data. If the theoretical results presented
here are accepted, then these data require a different, non-Griffiths
interpretation.  Ref. ~\cite{Neto98} argued that a disordered system
near a quantum critical point could be mapped onto the {\it disspationless}
Ising model in a transverse field; the results of the present
paper and of \cite{Millis01a} indicate on the contrary that dissipation
is essential. 

Ref.~\cite{Neto01} uses a novel variant of a technique introduced by Dotsenko
\cite{Dotsenko99} to study essentially the same model as is studied here. A
rather different result was obtained, namely that quantum Griffiths effects can
be important in a reasonable range of the phase diagram even in the metallic
case. We outline the differences between the results found here and those of
Ref.~\cite{Neto01}.
The method introduced by Dotsenko \cite{Dotsenko99} and used
by Ref \cite{Neto01} begins from a {\it classical}
theory defined by a functional integral with action given by
the static term in
Eq. \ref{S} and evaluates the disorder-average by the replica method.
Whereas other workers \cite{Boyanovsky84,Nayaranan01} then used the
replicated field theory to derive scaling equations for variables including
the mean disorder strength, Dotsenko argued that one should look for spatially
localized energetically {\it unstable}
configurations of the replicated field theory, which correspond to
local maxima of the replicated action and are to be identified with the
'droplets' discussed above. Dotsenko shows that the leading nonanalytic
contribution to the free energy in the vicinity of an assumed $T>0$ critical
point comes from droplets with size of the order of the magnetic correlation
length, $\xi$; we refer to these henceforth as 'typical droplets'. The authors
of Ref.~\cite{Neto01} assume that the $T\rightarrow0$ limit of this classical
theory may be straightforwardly taken, and then add to this theory estimates
of the dynamics of 'typical'
droplets. The results reported in Ref.~\cite{Neto01} disagree in a
number of specific details with the results presented here,
including for example the way in which the bare tunnelling rate
is estimated. The most important
difference, however, is in the interpretation of the results.
Ref.~\cite{Neto01} argues that one should identify the boundary of the
Griffiths region with the value of $\xi^{-2}$ at which a 'typical droplet'
ceases to tunnel. Our analysis, which involves averaging over all droplets,
indicates that independent of whether the 'typical droplet' (however defined)
may tunnel, the susceptibility is dominated by droplets which are at or beyond
the edge of ceasing to tunnel; these give an essentially 'superparamagnetic'
($\chi\sim1/T$) behavior, instead of the continuously varying exponent
characteristic of quantum Griffiths behavior. 

Ref. \cite{Neto00} presented a  detailed 
analysis of a different model in which spins are added to a pure system
which itself is far from any critical point.  
In this model
the phase transition is {\it disorder-driven}: it occurs
when the density of added spins is high enough that these
order; whereas our interest here has been in models in
which even the non-disordered system is near a critical point. 
Further in the model studied in ~\cite{Neto00} the way the
disorder is introduced means that 
the local spin amplitude $\phi_0$
(c.f our Eq. \ref{phi3d}) is always of order unity, whereas
in our treatment the local spin amplitude may be considerably smaller.  
An approximate
mapping between the model considered in \cite{Neto00} and the one considered
here may be obtained by setting our parameters $\phi_0$ and $\xi$ equal
to unity and 
considering the behavior as the disorder strength $V_0$ is increased
(whereas we consider a fixed $V_0$ and study the behavior as $\xi$ is increased).
 
Although specific details differ, in a broad qualitative sense
results obtained in Ref \cite{Neto00} are similar to those obtained here.
In particular, Ref \cite{Neto00} states that at sufficiently
low temperature dissipation will
suppress the quantum griffiths behavior. However, Ref
\cite{Neto00} argued that in an extremely wide temperature
regime could exist
in which behavior characteristic of the undamped system occurs,
whereas in the model we consider, for any reasonable parameters
there is no such temperature regime.  
A crucial point is that  \cite{Neto00} focussed
on model  parameters
such that the damping coefficient was extremely weak
(i.e in our notations (see below Eq. \ref{gamma})
they took  $c_{\gamma}<<1$). In this limit, it is plausible that 
there is  a temperature regime  in which behavior characteristic
of the undamped model may occur, before finally a crossover occurs 
to a regime (similar
to the one we considered) in which damping is important. Important
avenues for future investigation include more detailed studies of the crossovers
between the weak-damping and order unity damping cases and between
the disorder driven-criticality effects studied in \cite{Neto00} and the pure system
criticality-driven effects studied here, as well as determination of
the damping coefficient values appropriate to the heavy fermion materials
of interest. 

Our work has the following implications for experiment. First, the canonical
quantum Griffiths effects are due to weak disorder added to a pure critical
point. We have shown that in the limit of weak disorder and a pure critical
point described by the Hertz theory \cite{Hertz74,Millis95}, the dissipation
characteristic of metallic systems changes the quantum Griffiths singularities
into a kind of superparamagnetic behavior. In other words, as a matter of
principle the canonically defined quantum Griffiths behavior should
\textit{not} be observable in metals near magnetic quantum critical points.
This suggests that claims \cite{Maple98} to have observed quantum Griffiths
behavior in heavy fermion systems should be treated with caution (at least for
systems with Ising symmetry). Further, we showed that the droplets which
dominate the susceptibility can tunnel only when the system is not close to
criticality, and in these cases the droplet size is not much larger than the
basic scale of the theory. Thus if the susceptibility is
dominated by the tunnelling of droplets,
the picture which emerges is more
similar to the 'Kondo disorder' picture of \cite{Bernal95,Dobrosavljevic97} 
than it is to the conventionally-defined
quantum Griffiths picture. Indeed, the experimental claims involve 'heavy
fermion' systems where the interaction which favors a non-magnetic phase is
the Kondo effect. As noted by many authors \cite{Dobrosavljevic97}, the fact
that Kondo temperatures are exponentially sensitive to system parameters means
that a slight variation in system parameters can lead to a wide variation in
Kondo temperatures. The canonical assumption of weak disorder which we and
others \cite{Thill99,Rieger96,Neto01,McCoy69} have made may not be valid
for these systems. The interplay between quantum criticality and a broad
distribution of disorder should be treatable by the methods introduced here,
and seems worth examining.

A second point is that the very slow dynamics of the droplets makes it 
much easier
for them to order. Further, in a metallic system the droplet-droplet
interactions are of long range (see, e.g. \cite{Narozhny01} for a discussion
in the context of the two dimensional metal insulator transition). For this
reason we expect that in the presence of disorder the actual phase transition
at which long ranged order sets in is an essentially classical affair, in
which droplets lock together when the temperature becomes lower than some
droplet-droplet coupling.

A third point, perhaps relevant beyond the present context, is that
(as seen for example in Eq \ref{Sinstz2}) dissipation can have
a crucial effect  on bare tunnelling rates: in the metallic problem
we considered the cruical impediment to tunnelling of a droplet was
found to be the viscosity of the medium, not the energy barrier
which had to be surmounted.

\textit{Acknowledgements:} We thank  A. Castro-Neto, B. A. Jones and H. Rieger
for very helpful
conversations. This work was supported by NSF-DMR-00081075 (AJM), the Ames
Laboratory, operated for the U.S. Department of Energy by Iowa State
University under Contract No. W-7405-Eng-82 \ and an award from the Research
Corporation (J. S.). We acknowledge the hospitality of the Aspen Center for
Physics, where part of this work was performed.


\begin{thebibliography}{99}                                                                                               %


\bibitem {Thill99}M. J. Thill and D. A. Huse, Physica \textbf{A214 }321 (1995).

\bibitem {McCoy69}B. M. McCoy, Phys. Rev. Lett. \textbf{23} 283-6 (1969).

\bibitem {Guo96}M. Guo, R. N. Bhatt, and D. A. Huse, Phys. Rev. B 54,
3336-3342 (1996)

\bibitem {Rieger96}H. Rieger and A. P. Young, Phys. Rev. B 54, 3328-3335 (1996).

\bibitem {Fisher98}D. S. Fisher, Phys. Rev. Lett. \textbf{69} 534 (1992) and
Phys. Rev. \textbf{B51} 6411 (1995).

\bibitem {Lohneysen96}H. v. L\"{o}hneysen, T. Pietrus, G. Portisch, H. G.
Schlager, A. Schr\"{o}der, M. Sieck, and T. Trappmann, Phys. Rev. Lett. 72,
3262-3265 (1994) .

\bibitem {Mackenzie01}S.A Grigera., R.S. Perry, A. J. Schofield, M. Chiao, S.
R. Julian, G. G. Lonzarich, S. I. Ikeda, Y. Maeno, A. J. Millis and A. P.
Mackenzie, Science \textbf{294 }329-32 (2001).

\bibitem {Lonzarich99}C. Pfleiderer, G. J. McMullan, S. R. Julian, and G. G.
Lonzarich, Phys. Rev. \textbf{B 55}, 8330-8338 (1997)

\bibitem {Maple98}M. C. de Andrade, R. Chau, R. P. Dickey, N. R. Dilley, E. J.
Freeman, D. A. Gajewski, M. B. Maple, R. Movshovich, A. H. Castro Neto, G.
Castilla, and B. A. Jones, Phys. Rev. Lett. \textbf{81}, 5620-5623 (1998)

\bibitem {Neto98}A. H. Castro Neto, G. Castilla, and B. A. Jones, Phys. Rev.
Lett. \textbf{81}, 3531-3534 (1998)

\bibitem {Neto00}A. H. Castro Neto and B. A. Jones, Phys. Rev. \textbf{B 62},
14975-15011 (2000)

\bibitem {Neto01}A H Castro-Neto and B. A. Jones, cond-mat/0106176.

\bibitem {Stewart01}G. R. Stewart, Rev. Mod. Phys. \textbf{73}, 797-855 (2001).

\bibitem{Sherrington75} D. Sherrington and K. Mihill, J. Phys.
(Paris) {\bf 35} Colloque {\it C4} 199-201 (1975).

\bibitem {Millis01a}A. J. Millis, D. K. Morr, and J. Schmalian, Phys. Rev.
Lett. \textbf{87}, 167202 (2001).


\bibitem {Nayaranan01}R. Narayanan, T. Vojta, D. Belitz and T. R. Kirkpatrick,
Phys Rev. Lett \textbf{82 }5132 (1999) and Phys. Rev. \textbf{B60} 10150
(1999); R. Narayanan and T. Vojta, Phys. Rev. B 63, 014405 (2001).


\bibitem {Leggett87}A. J. Leggett, S. Chakravarty, A. T. Dorsey, M. P. A.
Fisher, A. Garg, and W. Zwerger, Rev. Mod. Phys. \textbf{59}, 1-85 (1987).

\bibitem {Dotsenko99}V. Dotsenko, J. Phys. \textbf{A32} 2949-60 (1999).

\bibitem {Boyanovsky84}D. Boyanovsky and J. L. Cardy, Phys. Rev. \textbf{B26},
154-170 (1982)



\bibitem {Hertz74}J. A. Hertz Phys. Rev. \textbf{B14}, 1165-1184 (1976)

\bibitem {Millis95}A. J. Millis, Phys. Rev. \textbf{B48}, 7183-7196 (1993).

\bibitem{Bernal95} O. O. Bernal, D. E. MacLaughlin, H. G. Lukefahr and B. Andraka,
Phys. Rev. Lett. {\bf 75} 2023-6 (1995).

\bibitem {Dobrosavljevic97} E. Miranda, V. Dobrosavljevic and B. G. Kotliar,
Phys. Rev. Lett. {\bf 78}  290-3 (1997) and E. Miranda and V. Dobrosavljevic,
Phys. Rev. Lett. {\bf 86} 264-7 (2001).

\bibitem {Narozhny01}B. N. Narozhny, I. L. Aleiner, and A. I. Larkin, Phys.
Rev. \textbf{B62}, 14898-14911 (2000)
\end{thebibliography}
\end{document}